\documentclass[aps,twocolumn,10pt,
superscriptaddress,
footinbib,
noeprint,
longbibliography,
prb]{revtex4-2}

\usepackage{graphicx}
\usepackage{dcolumn}
\usepackage{bm}
\usepackage{graphics}
\usepackage{epsfig,color}
\usepackage{physics2}
\usephysicsmodule{braket}
\usepackage{dsfont}
\usepackage{mathtools,amssymb}
\usepackage{multirow}
\usepackage{verbatim}
\usepackage{url}
\newcommand{\Tr}{\operatorname{Tr}}

\usepackage[breaklinks,colorlinks,bookmarks=false,citecolor=red,linkcolor=blue,urlcolor=blue]{hyperref}

\makeatletter
\newcommand\thefontsize{The current font size is: \f@size pt}
\makeatother

\begin{document}

\title{Exploiting emergent symmetries in disorder-averaged quantum dynamics}

\author{Mirco Erpelding}
\email[]{mirco.erpelding@uni-jena.de}
\affiliation{Institute of Condensed Matter Theory and Optics, Friedrich-Schiller-University Jena, Max-Wien-Platz 1, 07743 Jena, Germany}

\author{Adrian Braemer}
\email[]{bradrian@gmail.com}
\affiliation{Physikalisches Institut, Universit\"at Heidelberg, Im Neuenheimer Feld 226, 69120 Heidelberg, Germany}

\author{Martin Gärttner}
\email[]{martin.gaerttner@uni-jena.de}
\affiliation{Institute of Condensed Matter Theory and Optics, Friedrich-Schiller-University Jena, Max-Wien-Platz 1, 07743 Jena, Germany}

\date{\today}

\begin{abstract}
Symmetries are a key tool in understanding quantum systems, and, among many other things, can be exploited to increase the efficiency of numerical simulations of quantum dynamics. 
Disordered systems usually feature reduced symmetries and additionally require averaging over many realizations, making their numerical study computationally demanding.
However, when studying quantities linear in the time-evolved state, i.e.\ expectation values of observables, one can apply the averaging procedure to the time evolution operator, resulting in an effective dynamical map, which restores symmetry at the level of superoperators.
In this work, we develop schemes for efficiently constructing symmetric sectors of the disorder-averaged dynamical map using short-time and weak-disorder expansions. To benchmark the method, we apply it to an Ising model with random all-to-all interactions in the presence of a transverse field.
After disorder averaging, this system becomes effectively permutation-invariant, and thus the size of the symmetric subspace scales polynomially in the number of spins allowing for the simulation of large systems.
\end{abstract}

\maketitle



\section{Introduction}
The dynamics of quantum many-body systems is notoriously challenging to solve by analytical and even numerical means because the exponential scaling of the dimension of the underlying Hilbert space with the number of particles $N$.
An exception is systems that display global symmetries, which can be exploited to simplify calculations. 
For example, in permutation-invariant systems, the dynamics can be modeled by restricting to the space of fully symmetric states, leading to a computational complexity scaling polynomial in $N$, and even allowing analytical solutions in some cases.
This has enabled a deeper understanding of quantum many-body phenomena, including superradiance in the Dicke model \cite{Dicke1954,Hepp1973} or squeezing \cite{Ma2011} and (dynamical) quantum phase transitions \cite{Dusuel2004,Latorre2005,Zhang2017,Muniz2020} in collective spin systems, and provides scalable benchmarks for quantum devices \cite{Chinnarasu2025}.

Systems with quenched disorder, i.e.\ randomness in the Hamiltonian parameters, usually feature reduced symmetry. 
This breaks integrability and impedes scalable numerical simulations, also because of the fact that costly averaging over disorder realizations needs to be performed.
The plethora of intriguing physical phenomena related to disorder, including localization effects \cite{andersonabsenceofdiffusion, Abanin2019}, glassy dynamics \cite{RevModPhys.58.801, Ritort01062003}, and constrained relaxation \cite{PhysRevLett.53.1244, PhysRevLett.53.958}, and its ubiquity in experiments owing to noise in control parameters motivate the development of analytical and numerical techniques that overcome this issue.

A key insight that we exploit in this work is that after disorder averaging, symmetries can be restored.
Thus, for describing disorder-averaged quantities, symmetries of the distribution from which the random parameters in the Hamiltonian are drawn may be exploited.
In models with all-to-all random couplings, such as the Sachdev-Ye-Kitaev model \cite{Sachdev1993,Kitaev2015,Maldacena2016}, or the Sherrington-Kirkpatrick (SK) model \cite{sherringtonSolvableModelSpinGlass1975}, the system is permutation invariant on \emph{average} leading to exact solubility of equilibrium states, which has enabled ground-breaking insights in high-energy and condensed matter physics.
The role of average symmetries has also been discussed in the context of topological phases \cite{Ringel2012,Mong2012,Fulga2014,deGroot2022,Ma2023,Ma2025,Lee2025} and in quantum field theory \cite{Antinucci2023}.

Formally, disorder averaging can be accounted for at the level of quantum states, adopting an ensemble description.
The dynamics of the resulting disorder averaged mixed state is governed by an effective dynamical map between the initial and the time-evolved state, which can be recast into a Lindblad master equation up to a certain evolution time, but requires a non-Markovian description beyond \cite{kropfEffectiveDynamicsDisordered2016, gneitingDisorderdressedQuantumEvolution2020}. This formalism has been used to study various disordered quantum systems, such as the Anderson model or transport in weakly disordered media \cite{gneitingIncoherentEnsembleDynamics2016,gneitingQuantumEvolutionDisordered2017,gneitingDisorderRobustEntanglementTransport2019,gestssonEquivalenceDynamicsDisordered2024,paredesExploitingQuantumParallelism2005} and has also been extended to open quantum systems \cite{kropfOpenSystemModel2017}. 

\begin{figure}[b]
    \centering
    \includegraphics{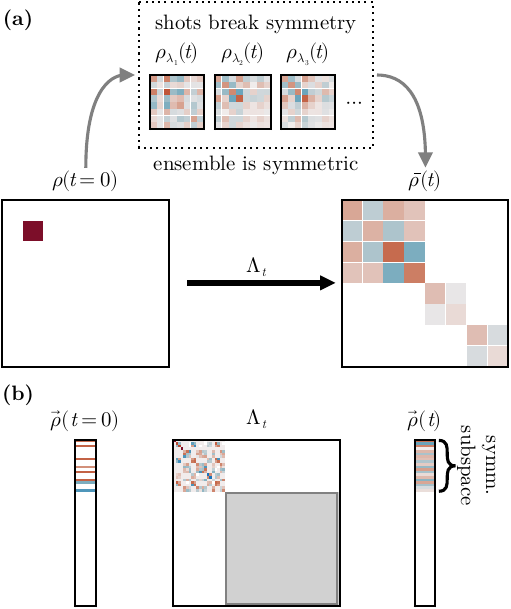}
    \caption{(a) The time evolution averaged over different realizations $\lambda_i$ of a disordered system, each with their respective states $\rho_{\lambda_i}$, can be understood as the action of an effective dynamical map $\Lambda_t$. If the ensemble of realizations is invariant under a symmetry (here permutations), then the average state is also compatible with the symmetry even though each shot by itself breaks the symmetry.
    (b) Representing the density operators as coefficient vectors with respect to an appropriate, i.e. symmetry-adapted, operator basis reveals the potential reduction in computational complexity arising because of the average symmetry. The dynamics remains confined within each symmetry sector, which is exponentially smaller than the full space in the case of average permutation invariance.}
    \label{fig:fig1}
\end{figure}

In this work, we exploit symmetries emerging under disorder averaging to construct an efficient numerical scheme for solving the dynamics of the disorder-averaged state. 
As illustrated in Fig.~\ref{fig:fig1}, the central idea is to construct the effective dynamical map $\Lambda_t$ directly within the relevant symmetry sector. 
In the case of average permutation invariance, this leads to a reduction of computational complexity to a polynomial scaling in $N$ \cite{Sandvik_2010, mansky2023permutationinvariantquantumcircuits}. 
As the construction of $\Lambda_t$ is not efficient in general, we resort to perturbative treatments, employing a short-time and a weak-disorder expansion.
We note that symmetrization is applied at the level of density matrices and superoperators, leading to an $N^3$ scaling of the dimension of the space of symmetrized states \cite{Sarkar1987,Hartmann2012,Xu2013}, in contrast to the linear scaling appearing in the Dicke \cite{Dicke1954} and Lipkin–Meshkov–Glick models \cite{Lipkin1965}, in which symmetrization is applied at the level of state vectors.
In case of the short-time expansion, we show that a Lindbladian description can be constructed, which provides a natural way of regularizing the perturbative series, allowing us to avoid unphysical long-time behavior. 
Moreover, we introduce a convenient mathematical formalism for treating the averaging process, leading to a straightforward derivation of weak-disorder expansions. 
We benchmark our approach by studying a transverse-field Ising model with independent, identically, and normally distributed all-to-all couplings. 
The utility of our techniques is further demonstrated in the context of the integrable SK model \cite{sherringtonSolvableModelSpinGlass1975}.

\section{Formalism}
\subsection{Effective disorder-averaged time evolution}
We consider systems in which disorder is introduced through evolution governed by different unitaries, each associated with a Hamiltonian $H_{\lambda}$ that depends on randomly varying parameters $\lambda$. We restrict our considerations to a finite number of fluctuating parameters, specifically $\lambda \in \mathds{R}^n$ for some appropriate $n\in \mathds{N}$. Each disorder configuration of the system, referred to as a shot, or random realization, is associated with a probability $p(\lambda)$. In many cases, one can transform the probability density $p(\lambda)\rightarrow p(\tilde{\lambda})$, such that $H_{\tilde{\lambda}}=\sum_{i}\tilde{\lambda}_{i}O_{i}$ with Hermitian operators $O_{i}$. Thus, we focus on Hamiltonians with disorder in their linear coefficients. Initially, the system is prepared in a state $\rho_0$ that is not necessarily pure. 
Averaging the expectation value of an observable $O$ over possible random realizations yields
\begin{align}
    \overline{\braket*[1]{O}(t)} &= \int\!\mathrm{d}p(\lambda)\ \Tr O U_\lambda(t) \rho_0 U_\lambda(t)^\dagger\\
    &= \Tr O \int\!\mathrm{d}p(\lambda)\ U_\lambda(t) \rho_0 U_\lambda(t)^\dagger\\
    &\equiv \Tr O \overline{\rho}(t),
\end{align}
where $\overline{\rho}(t)$ is the effective state resulting from the statistical mixture of different shots, where the overbar denotes the disorder averaging. The time evolution of this effective state is described by the dynamical map $\Lambda_t$ defined via
\begin{align}
    \overline{\rho}(t) &= \int\!\mathrm{d}p(\lambda)\ U_\lambda(t) \rho_0 U_\lambda(t)^\dagger \\
    &= \int\!\mathrm{d}p(\lambda)\ \exp\left(-it [H_\lambda, \cdot] \right) \rho_0\\
    &\equiv \int\!\mathrm{d}p(\lambda)\ \mathcal{U}_{\lambda}(t) [\rho_{0}]\\
    &\equiv \Lambda_t[\rho_0].
\end{align}
We refer the reader to Appendix \ref{app:SuOpFormalism} for details on the notation. Notably, even if the initial state is pure and each shot evolves unitarily, the effective state $\overline{\rho}(t)$ will be mixed because of the incoherent averaging over different unitary evolutions.

As discussed above, while global symmetries, such as spatial translation or reflection symmetries, are typically broken in each individual shot, they are reestablished under disorder averaging if the probability measure is invariant under a given symmetry transformation.
This directly implies that the dynamical map $\Lambda_t$ retains these symmetries. More formally, the ensemble of realizations must remain invariant under the group action of the corresponding symmetry transformations. That is, if $G$ is the associated group, $T:G\rightarrow U(\mathcal{H})$ its unitary representation on the Hilbert space $\mathcal{H}$, $dp(\lambda)=dp(g(\lambda))$ for all $g \in G$ and for all $H_{\lambda}$ in the ensemble $T_{g}H_{\lambda}T_{g}^{\dagger}=H_{g(\lambda)}$ is in the ensemble as well, then any symmetry transformation in $G$ commutes with the effective dynamical map:
\begin{align}
    T_{g}\bar{\rho}T_{g}^{\dagger}&=\int dp(\lambda)\ T_{g}U_{\lambda}(t)\rho_{0}U_{\lambda}(t)^{\dagger}T_{g}^{\dagger}\\
    &=\int dp(\lambda)\ U_{g(\lambda)}(t)T_{g}\rho_{0}T_{g}^{\dagger}U_{g(\lambda)}(t)^{\dagger}\\
    &=\int dp(g(\lambda))\ U_{g(\lambda)}(t)T_{g}\rho_{0}T_{g}^{\dagger}U_{g(\lambda)}(t)^{\dagger}\\
    &=\Lambda_{t}[T_{g}\rho_{0}T_{g}^{\dagger}].
\end{align}
Consequently, if the initial state is invariant under a given symmetry transformation, then $\overline{\rho}(t)$ will retain this symmetry at all times. For example, if the initial state is permutation-invariant and the ensemble of Hamiltonians $H_{\lambda}$ is invariant under permutations of particle labels, $\overline{\rho}(t)$ will stay within the subspace permutation-invariant states. Consequently, to evaluate its dynamics, we need only consider the relevant part of $\Lambda_t$ that couples states of the relevant symmetry sector, which can significantly simplify numerical calculations, as illustrated in Fig.~\ref{fig:fig1}.

\subsection{Effective Lindbladian description}
Kropf \textit{et al.} \cite{kropfEffectiveDynamicsDisordered2016} derived a description of the disorder-averaged dynamics in terms of a Lindblad Master Equation using the effective dynamical map. Assuming invertibility of $\Lambda_{t}$, the effective Lindbladian can be defined as
\begin{align} \label{eq.:Lindblad_defining_property1}
\mathcal{L}_{t}&=\dot{\Lambda}_{t}\circ\Lambda_{t}^{-1},\\\notag
\dot{\bar{\rho}}(t)&=\mathcal{L}_{t}[\bar{\rho}(t)]\\ \label{eq.:Lindblad_defining_property2}
&=-i[H(t),\bar{\rho}(t)]\\\notag
&+\sum_{k}\gamma_{k}(t)\left(L_{k}(t)\bar{\rho}(t)L_{k}^{\dagger}(t)-\frac{1}{2}\left\{L_{k}^{\dagger}(t)L_{k}(t),\bar{\rho}\right\}\right), 
\end{align}
resulting in a time-local master equation with time-dependent jump operators $L_{k}(t)$, rates $\gamma_{k}(t)$ and effective Hamiltonian $H(t)$.
As a generator of time evolution, the Lindbladian is confined to the same symmetry sectors as the dynamical map. This approach may provide the advantage that the effects of disorder averaging become physically interpretable as dephasing and dissipation.
However, computing the relevant part of $\Lambda_t$ or $\mathcal{L}_{t}$ --- that is, representing it in a single symmetry sector --- is generally inefficient for non integrable models. 

The main contribution of our work is to show that the effective dynamical map can be directly and efficiently constructed within the relevant symmetry sector when perturbatively expanding around an analytically tractable point. Specifically, we present two different approaches for treating short-time and weak-disorder expansions, respectively.
This task is made significantly easier by employing differential operations to avoid explicit averaging over the disorder distribution. Specifically, the disorder average of a quantity $f(\lambda)$ can be represented as 
\begin{align}
\overline{f}=\left.\phi(-i\nabla_{\mu})f(\mu)\right|_{\mu=0}\equiv Df,
\end{align}
where $\phi(k)$ is the characteristic function of the disorder distribution, defined via
\begin{align}
    p(\lambda)=\int \frac{d^nk}{(2\pi)^{\frac{n}{2}}}\ \phi(k)\ e^{-ik\lambda}.
\end{align}
For details on the derivation see Appendix \ref{app:Averaging_Derivative}. Acting with $D$ on $\mathcal{U}$ by expanding $\phi(-i\partial_{\mu_{1}},\dots,-i\partial_{\mu_{n}})$  into a formal power series, we obtain an expansion of the dynamical map into symmetric operators, which can be extracted order by order. The Lindbladian takes the form of a projected partial differential equation (see Appendix \ref{app:PDE_perspective}). 
This approach allows for the straightforward solution of integrable systems, as exemplified by the Sherrington-Kirkpatrick model in Appendix \ref{app:exampleSKModel}.

\subsection{Regulating expansions}
Truncating the time expansion of the dynamical map to an order $k\in\mathds{N}$ will lead to a corresponding polynomial behavior in $t$, resulting in divergent expectation values. In contrast, the association with an open quantum system suggests some form of decay. Regularization allows us to adapt higher-order terms to prevent divergence in time without altering the approximation order. If $f\in C^{\infty}(\mathds{R})$ is an invertible function, we can rewrite $\Lambda_{t}=f\circ f^{-1}(\Lambda_{t})$ and subsequently expand $f^{-1}(\Lambda_{t})$. The truncation will then be regulated by $f$.  Therefore, the choice of $f$ must be made with care. A reasonable choice also necessitates caution regarding the truncation order itself, as the asymptotic behavior heavily depends on the sign of the highest order term. 

In Sec.~\ref{sec:applications}, we will discuss several possible choices of regularizing functions $f$ along with the employed short-time and weak-disorder expansion methods. We will recover the short-time expansion of the Lindbladian as one natural example for such a regularization scheme.

\section{Applications}
\label{sec:applications}
We demonstrate the power of our approach by considering the transverse-field Ising model 
\begin{align}
\label{eq.:Hamiltonian}
    H=\sum_{i<j}\frac{J_{ij}}{\sqrt{N}}Z_{i}Z_{j}+h\sum_{i}X_{i}
\end{align}
with randomly distributed all-to-all couplings $J_{ij}$, independently drawn from a normal distribution $\mathcal{N}( \bar{J}, \sigma)$. $h$ is the strength of the transverse field, which we set to $h = 1$ in our numerical simulations, and the number
of particles is denoted by $N$. Under disorder averaging, this model features an effective permutation symmetry. Therefore, an initial state that is permutation invariant, such as a fully polarized state, where all spins are in the same state, will remain permutation invariant on average. We can thus represent the time-evolved state and the effective dynamical map efficiently in a symmetrized Pauli string basis. We denote the symmetric string, which is normalized with respect to the Frobenius norm, as $\Sigma_{(x,y,z)}$, because it only depends on the number of $x$-, $y$- and $z$-Pauli matrices it contains. For details on this notation, see Appendix \ref{app:symmetric_pauli_strings}. If not otherwise stated, we will use the $Z$-polarized initial state $\rho_{0}=\ket{\uparrow\dots\uparrow}\bra{\uparrow\dots\uparrow}=2^{-N/2}\sum_{i=0}^N \Sigma_{(0,0,i)}$, for our calculations.

\subsection{Short-time expansion}
The most straightforward example of a perturbative treatment is a short-time expansion. However, expanding the dynamical map $\Lambda_{t}$ to low orders is insufficient to accurately reflect physical effects, as truncations of the expansion will ultimately lead to unboundedness of the approximated map. Consequently, the trace norm could exceed the boundaries of physical states, i.e. $\lVert\rho\rVert\leq1$. Therefore, the main idea of this section is to perform a short-time expansion of the effective Lindbladian instead of $\Lambda_{t}$, allowing us to avoid issues of unboundedness and thus non-physicality. This formally corresponds to choosing the time-ordered exponential 
\begin{align}
    \label{eq:timeorderdexponential_Lambda_t}
    \Lambda_{t}=\mathcal{T}\exp{\left(\int_{0}^{t}\mathcal{L}_{t'}\ dt'\right)}
\end{align}
as the regularizing map $f$ with $f^{-1}(\Lambda_t)=\mathcal{L}_t
$. Regularization is achieved by applying $f$, i.e. solving the  Lindblad master equation.
Using the defining property \eqref{eq.:Lindblad_defining_property1} of $\mathcal{L}_{t}$ and the Leibniz rule, we find a recursive expansion formula 
\begin{align}
    \notag
    &\mathcal{L}_{t}&=&\ \sum_{n=0}^{\infty}\frac{t^n}{n!} \mathcal{L}^{(n)}_{t=0}\\\notag
    &\mathcal{L}^{(n)}_{t=0}&=&\ (-i)^{n+1}\overline{[H_{\lambda},\ \cdot\ ]^{n+1}}\\
    &&&-\sum_{j=0}^{n-1}(-i)^{n-j}\binom{n}{j}\mathcal{L}^{(j)}_{t=0}\overline{[H_{\lambda},\ \cdot\ ]^{n-j}}. \label{eq.:Lindblad_time_expansion}
\end{align}
Equation \eqref{eq.:Lindblad_time_expansion} allows a natural definition of generalized cumulants given by
\begin{align}
    \notag
    &\kappa_{n}\equiv i^{n}\mathcal{L}^{(n-1)}_{t=0},\ \text{such that}\\ \label{eq.:Lindblad_cumulants}
    &\kappa_{n}=\overline{[H_{\lambda},\ \cdot\ ]^{n}}-\sum_{j=1}^{n-1}\binom{n-1}{j-1}\kappa_{j}\overline{[H_{\lambda},\ \cdot\ ]^{n-j}}.
\end{align}
For each order, the cumulants can be explicitly calculated in the symmetric subspace (cf. Appendix \ref{app:Lindbladterms}) because each moment of the Hamiltonian's commutator is a superoperator, which preserves the symmetric space. Furthermore, Eqs.~\eqref{eq.:Lindblad_time_expansion} and \eqref{eq.:Lindblad_cumulants} imply that in the limit of no disorder $p(\lambda)\rightarrow \delta{(\lambda -\overline{\lambda})}$, only $\mathcal{O}(t^0)$ corresponding to $i\left[\ \cdot\ ,\bar{H}\right]$ remains because higher-order moments simply factorize leading to order-wise cancellation. Thus, the short-time expansion of the Lindbladian is already exact at zeroth order in the disorder-free limit and recreates the von Neumann equation with the average Hamiltonian. It is also easy to see from Eqs.~\eqref{eq.:Lindblad_time_expansion} and \eqref{eq.:Lindblad_cumulants} that every even order cumulant corresponds to a dephasing operator and can be cast into jump-operator form. For example, see the Lindbladian of the SK-model in Appendix \ref{app:exampleSKModel} which is the same for the transverse field model up to linear order in time (cf. Appendix \ref{app:Lindbladterms}) or compare with the short-time
approximation in \cite{kropfEffectiveDynamicsDisordered2016} that we recover by truncation of Eq.~\eqref{eq.:Lindblad_time_expansion} at
order $\mathcal{O}(t^2)$. Every odd cumulant yields a contribution to the effective unitary dynamics.

\begin{figure}[b]
    \centering
    \includegraphics[width=0.468\textwidth]{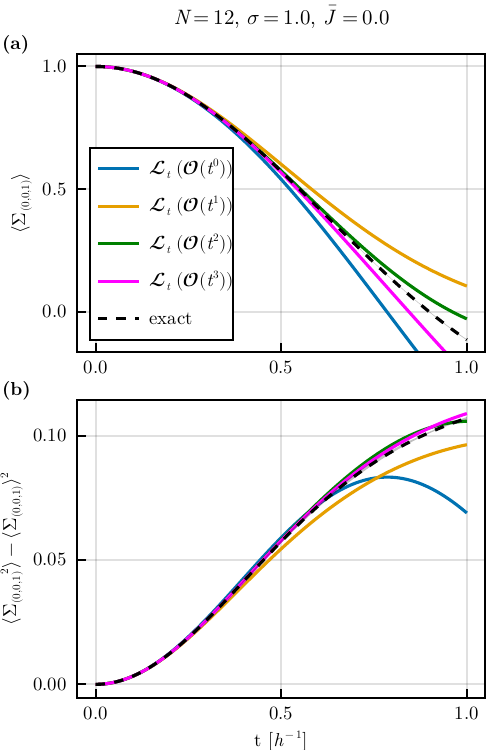}
    \caption{Time expansion ansatz. Solving Eq. \eqref{eq.:Lindblad_defining_property2} for $\rho_{0}=\ket{\uparrow\dots\uparrow}\bra{\uparrow\dots\uparrow}$ and $\mathcal{L}_{t}$ expanded up to $\mathcal{O}(t^3)$. Exact results have been obtained averaging over $1000$ disorder shots. The corresponding statistical error is indicated by the width of the gray line (smaller than the default line width here).  Panel (a) shows the average total $Z$-magnetization $\langle \Sigma_{(0,0,1)}\rangle$ over time in units of $h^{-1}$. Panel (b) shows its variance. In the regime $t\ll\epsilon_{max}^{-1}$ the accuracy increases systematically with the approximation order.}
    \label{fig:lindblad_expansion_orders}
\end{figure}

\begin{figure*}
    \centering
    \includegraphics[width =1\textwidth]{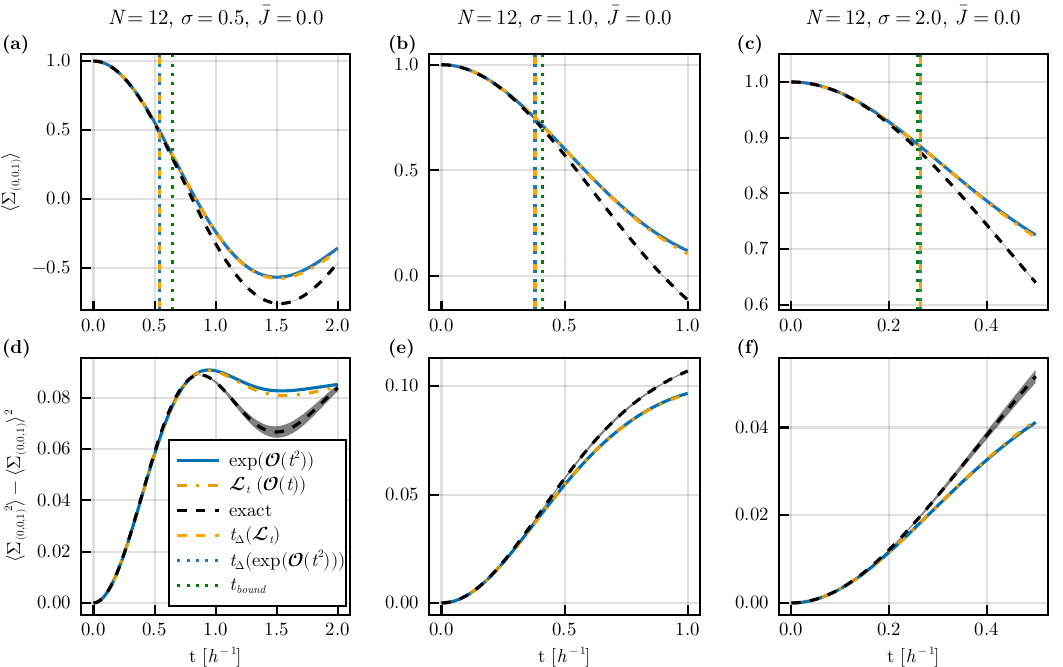}
    \caption{Short-time expansion ansatz. Solving Eq. \eqref{eq.:Lindblad_defining_property2} for an expanded Lindbladian truncated after order $\mathcal{O}(t)$. Simultaneously the exponential regularized expansion of the dynamical map up to order $\mathcal{O}(t^2)$ (blue) and exact results (black) are plotted against time in units of $h^{-1}$ for a vanishing mean coupling strength $\bar{J}=0$. Exact results were obtained using 1000 disorder shots resulting in noticeable statistical fluctuations (gray) in the variance.  Panels (a)-(c) show the average total $Z$ magnetization $\langle \Sigma_{(0,0,1)}\rangle$ for different disorder strengths $\sigma$, while in panels (d)-(f) the variances in $Z$ are displayed. First deviations of the approximations from the exact results on the order of the corresponding variance are marked and compared to the estimate region $t_{bound}$ proving reliability of the expansion in predictable regimes (vertical lines).} 
    \label{fig:time_expansion_sigmas}
\end{figure*} 

Truncating the short-time expansion at $\mathcal{O}(t^k)$ results in an error of order $\mathcal{O}(t^{k+2})$ in the states or observables, as is clear from Eq.~\eqref{eq:timeorderdexponential_Lambda_t}. 

We emphasize that alternative choices of regularizing functions may have practical advantages, as they can reduce the number of required terms to be evaluated even further. For instance, instead of evaluating each moment $\overline{[H_{\lambda},\ \cdot\ ]^{n}}$, the exponential regularization --- that is, using the regularization function $f(x)=\exp(x)$ --- requires only specific operator combinations, which may facilitate projection onto the symmetric subspace. In turn, one has to be careful regarding the convergence of the series. The Taylor series of the inverse maps of $\exp(x)$ and $x^{-1}$ [the second one is used to define $\mathcal{L}_t$, cf. Eq.~\eqref{eq.:Lindblad_defining_property1}] both have a convergence radius of one, while the largest singular value of the effective dynamical map is smaller or equal to one. Therefore, the series resulting from expanding the inverse map needs to be checked for convergence. In many cases the series may become asymptotic; for example the series of the Lindblad expansion fails if $\Lambda_t$ is not invertible anymore. If it converges, we expect the approximation to be valid for $t\ll\epsilon_{max}^{-1}$, with $\epsilon_{max}=\max\{\bar{J},h,\sigma\}$, the largest energy scale of the problem.

Figure \ref{fig:lindblad_expansion_orders} illustrates the convergence of the Lindblad solution to the exact one with increasing truncation order in this regime: A systematic improvement can be seen up to $ht\approx 0.5$. Furthermore, even the late-time behavior seems to improve slightly every second order despite convergence not being guaranteed for $ht>0.5$. 

In the regime where invertibility of $\Lambda_{t}$ is given, we can get a tighter error bound using the leading-order error term, i.e.\ the next-order term of the expansion. Figure~\ref{fig:time_expansion_sigmas} shows the Lindblad solution truncated after linear order, along with an exponential regularization approach -- expanding $\log(\Lambda_{t})$ and inserting it into an exponential after truncation -- that is truncated after second order for different disorder strengths. 
Note that these truncation orders are equivalent, since Eq.~\eqref{eq:timeorderdexponential_Lambda_t}
shows that truncation of $\mathcal{L}_{t}$ at $\mathcal{O}(t^k)$ corresponds to truncation of $\int \mathcal{L}_{t}\ dt$ at $\mathcal{O}(t^{k+1})$. For convergence we require $||\int_{0}^{t}dt'\ (t'^2/2)\mathcal{L}_{t=0}^{(2)}||=||(t^3/3)\mathcal{L}_{t=0}^{(2)}||\ll1$, which implies 
$t\ll\sqrt[3]{3/(4\sigma^{2}(N-1))}\equiv t_{bound}$.

To benchmark the advantage of the new bound over the naively expected range of agreement, we calculated the first moment of the $Z$-magnetization and its variance. Together with $t_{bound}$ we define $t_{\Delta}$ as the first point in time at which the deviation of the approximation from the exact first moment of the magnetization is of the same order of magnitude as the variance.

Figure \ref{fig:time_expansion_sigmas}  illustrates that the expansion closely approximates the behavior over a surprisingly wide range compared to $t_{bound}$, which establishes it as a reliable upper bound on the regime in which the approximation error remains controlled (given that the expansion exists). The choice of regularization is, on these scales, nearly irrelevant, as evidenced by the matching $t_{\Delta}$ marks. Therefore, the selected regularization should prioritize numerical efficiency. Thus, choosing the exponential regularization over the Lindblad expansion is recommended for this model. 

\begin{figure*}
    \centering
    \includegraphics[width = 1\textwidth]{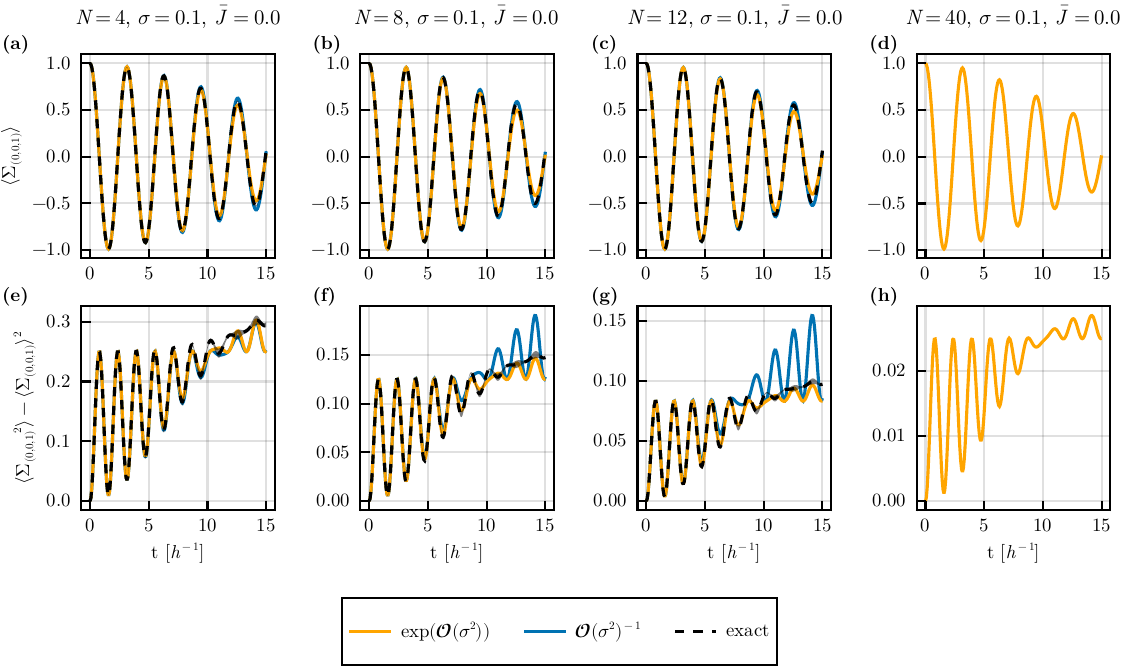}
    \caption{Weak-disorder expansion of transverse-field Ising model. Exponential (orange) and inverse (blue) regularization schemes are plotted together with exact calculations (black) for different system sizes. Panels (a)-(c) show the total $Z$ magnetization for $N=4,\ 8$ and $12$, while (e)-(g) illustrate the corresponding variances. In (d) and (h) results for $N=40$ are given for the exponential regularized weak-disorder expansion (blue). The deviation at $t\approx\sigma^{-1}$ is clearly observable, showcasing the power of the exponential regularization.}
    \label{fig:weak_disorder_expansionN12}
\end{figure*}

\subsection{Weak-disorder expansion}
\label{sec:weak_disorder_expansion}
In the case of independent normally distributed couplings $J_{ij}$, the characteristic function 
is given by $\phi(k)\propto \exp{\{ik\cdot\bar\lambda-\sigma^2 k^2/2\}}$ up to a normalization factor. The term $\exp{(i(-i\nabla_{\mu})\bar\lambda)}$ of the exponential in $\phi(-i\nabla_{\mu})$ is applied by setting $\mu=\bar\lambda$ instead of $\mu=0$ after differentiation. Note that this is a translation of the auxiliary variable $\mu$ to the mean $\bar\lambda$, an explicit calculation is in Eq.~\eqref{eq:WDE_translation}. Then, the averaging operator to deal with is $D_{\mu}\equiv\exp{\{-\sigma^2/2 \ \triangle_{\mu}\}}$.

At weak-disorder $\sigma^2 \ll1$ we can expand the dynamical map as a power series in $\sigma^2$, which from the discussion above corresponds to an expansion of $D_{\mu}$ in $\sigma^2$ acting on $\mathcal{U}_{\mu}(t)$. 
We exploit the fact that the average Hamiltonian corresponds to an integrable system to analytically calculate the derivative terms appearing in the expansion coefficients (see Appendix \ref{app:derivation_weak_time_evolution} for details). To first order in $\sigma^2$ we find
\begin{align}
\label{eq:WDE_Lambda_t}
\notag    \Lambda_{t}=&\phantom{-}\bar{\mathcal{U}}_{t}\left\{\mathds{1}-\frac{\sigma^2}{N}\sum_{i<j}\int_0^{t}\!\!dt'\! \int_{0}^{t'}\!\!d\tilde{t}\ [Z_iZ_j(t'),[Z_iZ_j(\tilde{t}),\:\cdot\:]]\right\}\\
    &+\mathcal{O}(\sigma^{4})\\
    =&\phantom{-}\bar{\mathcal{U}}_{t}\left\{\mathds{1}-\sigma^2 O(t)\right\}+\mathcal{O}(\sigma^{4}),
\end{align}
where $Z_{i}Z_{j}(t)$ are Heisenberg operators of $Z_{i}Z_{j}$ evolving under $\bar{H}$ and $\bar{\mathcal{U}}_t$ its corresponding time-evolution superoperator. 
The resulting leading-order term scales as $(\sigma t)^2$, which leads to a $t^2$ dependence of observables, as for the time expansion ansatz. With our initial assumption of weak disorder, i.e.\ $\sigma^2\ll1$, we may interpret the expression in brackets as either an exponential 
\begin{align*}
    \Lambda_{t}\approx&\phantom{-}\bar{\mathcal{U}}_{t}\exp\left\{-\sigma^2 O(t)\right\}
\end{align*}
or as the leading order expansion of an inverse operator 
\begin{align*}
    \Lambda_{t}\approx&\phantom{-}\bar{\mathcal{U}}_{t}\left\{\mathds{1}+\sigma^2O(t)\right\}^{-1}.
\end{align*}
Both approximations control the late-time behavior and thus we refer to them as exponential or inverse regularization. The exponential regularization corresponds to the solution of Eq. (5) in \cite{gneitingDisorderdressedQuantumEvolution2020}, which underlines the strength of our method as a versatile and efficient means of deriving disorder averaged effective dynamics.

Figure \ref{fig:weak_disorder_expansionN12} shows the $Z$-magnetization and its variance for an initially $Z$-polarized state, calculated using both exponential and inverse regularization methods across different spin numbers $N$. Note that the SK contribution of the Hamiltonian in Eq.~\eqref{eq.:Hamiltonian} is scaled by the factor $1/\sqrt{N}$ to achieve $N$-independence of relevant time scales. The magnetization and its variance exhibit periodic behavior, where the oscillation is a simple Rabi oscillation due to $\bar{H}$ and the decrease, or respectively, increase, of the envelope curve is caused by disorder-induced decoherence. Furthermore, the variance shows an expected $1/N$-scaling. All these features are well captured by exponential and inverse regularization; however, they begin to deviate significantly between $t=10 h^{-1}$ and $t=7.5h^{-1}$, which is better than the naively expected regime of $t\ll\sigma^{-1}=10h^{-1}$. Beyond this range, the quality of the approximation varies between the different regularization schemes employed. Exponential regularization extends the range of agreement between prediction and exact solutions to approximately $t<10h^{-1}=\sigma^{-1}$, while the inverse regularization shows earlier deviations as $N$ increases. Additionally, for the inverse regularization  the deviations of the magnetization variance become stronger and start earlier for increasing $N$. Thus, the exponential approach is preferable.
The domain in which the exponential-regularized scheme is in good agreement with exact simulations, is $N$ independent. Thus, we can apply our method to compute the dynamics for larger systems sizes, beyond the reach of exact simulations, as demonstrated in Figs.~\ref{fig:weak_disorder_expansionN12}(d) and \ref{fig:weak_disorder_expansionN12}(h). 

Errors arise at $\mathcal{O}(t^3)$, as indicated by the derivation in Appendix \ref{app:derivation_weak_time_evolution}. In contrast to time expansion, the relevant energy scale depends on $\sigma/h$, which means that $\sigma$ needs to be small relative to other energy scales. Thus, the main advantage over time expansion is that larger time scales are captured more accurately .

We also studied the time-averaged late-time values of the magnetization (diagonal ensemble values) for the mixed-field Ising model (see Appendix \ref{app:X_magnetization}). We observe that the weak-disorder expansion perfectly matches the exact solution for a wide range of initial states and field-configurations, calling for a deeper analytical understanding.

\section{Conclusions}


By exploiting symmetries that are broken for individual disorder realizations but hold on the level of the entire disorder ensemble, we constructed efficient perturbative schemes for simulating the disorder-averaged dynamics of quantum many-body systems.
The construction of the resulting effective dynamical map directly within the relevant symmetry sector is efficiently possible order-by-order in short-time or weak-disorder expansion schemes.
Our numerical results show that, while our expansion schemes are only reliable within a certain convergence radius, choosing a suitable regularization scheme, allows one to obtain good results even beyond this point, and that, in some cases, the asymptotic late-time-values are predicted correctly.

We emphasize that our technique is widely applicable, as it does not depend on a particular choice of model or disorder distribution. In fact, for the weak-disorder expansion even the existence of an analytically solvable point to expand around is not required and could be worked around numerically. Thus, there is a wide range of interesting applications of our technique in quantum simulation experiments with inherent positional randomness such as cold atomic gases or color centers in diamond \cite{Yan2013,kucskoCriticalThermalizationDisordered2018,signolesGlassyDynamicsDisordered2021}. 
Since in these systems one cannot assign consistent labels to individual spins across shots, any accessible observable is naturally permutation-invariant. Thus, our method could enable scalable numerical simulations in this area serving as benchmarks for experiments and potentially leading to new discoveries.

The approximations and regularizations presented here represent but a first step in the direction of developing efficient numerical tools for simulating disorder-averaged dynamics. An example of further methodological developments that we foresee is sketched out in Appendix~\ref{app:PDE_perspective} where we derive a partial differential equation on the product space of superoperators and polynomials that is equivalent to the effective Lindbladian and might offer different ways to approximate the dynamics. Finally, we remark on the conceptual similarity of our formalism to the path-integral formulations of Quantum Field Theory (QFT). While in QFT the quantum effective action arises from the coherent sum over different histories, in our scenario the incoherent sum over disorder realizations gives rise to an effective Lindbladian.
Thus, it might be possible to exploit this analogy to transfer computational techniques from the realm of Quantum Fields to the study of disordered quantum systems, e.g.\ by using renormalization group methods established in QFT to develop new tools for efficient computations of effective dynamics.

\section*{Acknowledgments}
This research is supported by funding from the German Research Foundation (DFG) under the Project Id 398816777-SFB 1375 (NOA). A.B. acknowledges support
by the Deutsche Forschungsgemeinschaft (DFG, German Research Foundation) within the Collaborative Research Center SFB1225 (ISOQUANT) under the Project Id 273811115.
For the numerical work, we used the Julia programming language \cite{bezansonJuliaFreshApproach2017} and the following packages/tools: \texttt{DrWatson.jl}~\cite{datserisDrWatsonPerfectSidekick2020}, \texttt{Pluto.jl}~\cite{fonsvanderplasFonspPlutoJl2024}, \texttt{DifferentialEquations.jl}~\cite{rackauckas2017differentialequations} with Verner's integration schemes~\cite{vernerNumericallyOptimalRunge2010}, \texttt{Makie.jl}~\cite{danischMakieJlFlexible2021}.

M.E.\ developed the analytical methods and carried out the numerical computations under the supervision of A.B.\ and M.G. M.E.\ drafted the manuscript with contributions from A.B. All
authors contributed to the finalization of the manuscript.

\section*{Data Availability}
The data that support the findings of this article are openly
available \cite{myrepo}.

\appendix
\section{Superoperator formalism and notation}
\label{app:SuOpFormalism}
Time evolution in quantum mechanics is governed by a linear differential equation, and thus the time-evolved state is related to the initial state through a linear operator (time evolution operator). For time-independent closed systems, the time evolution operator is just the exponential of the Hamiltonian $U(t)=\exp(-iHt)$. When describing the evolution of general mixed quantum states, one uses the density operator formalism, in which unitary evolution reads $\rho(t) = U(t)\rho_0U(t)^\dagger$. For our application, a formulation in terms of a single linear operator acting from the left onto the state is preferable. Thus, we "vectorize" the density matrix leading to $\vec{\rho}(t) = U(t)\otimes U(t)^{*}\vec{\rho}_0$. Here we explain the mathematical background of this equivalent formulation and define a useful shorthand notation.

At the heart of this "vectorized" density matrix formalism is the isomorphism defined by $\Phi:\, A_{ij}e_{i}\otimes \epsilon_{j}\mapsto A_{ij}e_{i}\otimes e_{j}$ where $e_{i}$ is a basis vector and $\epsilon_{i}$ is its canonical dual. Let $A,\ B$ and $C$ be matrices and denote $\vec{C}=\Phi(C)$ then
\begin{align}
\notag
ACB&=A_{ij}C_{jl}B_{ls}e_{i}\otimes \epsilon_{s}\\\notag
&\simeq A_{ij}C_{jl}B_{ls}e_{i}\otimes e_{s}\\\notag
&=A_{ij}B^{T}_{sl}C_{jl}e_{i}\otimes e_{s}\\\notag
&=A_{ij}B^{T}_{sl}C_{kr}\epsilon_{j}(e_{k}) \epsilon_{l}(e_{r})e_{i}\otimes e_{s}\\\notag
&=A\otimes B^{T}\cdot \vec{C}.
\end{align}

For any $A,B\in \text{Mat}_{n}(\mathbb{C})$ the commutator is given by
\begin{align}
\notag
    \left[A,B\right]&=AB-BA=AB\mathds{1}-\mathds{1}BA\\\notag
    &\simeq A\otimes \mathds{1}^{T} \cdot \vec{B} - \mathds{1}\otimes A^{T} \cdot \vec{B}\\\notag
    &=(A\otimes \mathds{1}^{T}  - \mathds{1}\otimes A^{T}) \cdot \vec{B}\\
    \Rightarrow\left[A,\ \cdot\ \right]&\simeq A\otimes \mathds{1}  - \mathds{1}\otimes A^{T}.
\end{align}
The exponential map of matrices is defined by the formal power series
\begin{align}
    \exp{\{A\}}=\sum_{k=0}^{\infty}\frac{A^k}{k!},
\end{align}
which is absolute convergent and exists for any $A$. The tensor product of two matrix exponentials is then given by
\begin{align}
\notag
    \exp{\{A\}}\otimes\exp{\{B\}}&=\left[\exp{\{A\}}\otimes \mathds{1}\right]\left[\mathds{1}\otimes\exp{\{B\}}\right]\\\notag
    &=\left[\sum_{k=0}^{\infty}\frac{1}{k!}(A\otimes \mathds{1})^{k}\right]\left[\sum_{k=0}^{\infty}\frac{1}{k!}(\mathds{1}\otimes B)^{k}\right]\\\notag
    &=\exp{\{A\otimes \mathds{1}\}}\exp{\{\mathds{1}\otimes B\}}\\
    &=\exp{\{A\otimes \mathds{1}+\mathds{1}\otimes B\}},
\end{align}
where we used the Baker-Campbell-Hausdorff formula and $\left[a\otimes \mathds{1},\mathds{1}\otimes b\right]=0$ in the last step. Hence, the time-evolution operator acting on density matrices can be written as
\begin{align}
\notag
    \mathcal{U}(t)\rho_{0}&=U(t)\rho_{0}U(t)^{\dagger}\simeq U(t)\otimes U(t)^{*} \vec{\rho}_{0}\\\notag
    &=\exp{\{-it (H\otimes \mathds{1}- \mathds{1} \otimes H^{T})\}} \vec{\rho}_{0}\\
    &\simeq\exp{\{-it[H,\ \cdot\ ]\}}\rho_{0}.
\end{align}
For simplicity of notation, we suppress the arrow over the vectorized density matrix, since it is clear which representation to use by the representation of the operators acting on it.

\section{Distribution Average as Differential Operator}
\label{app:Averaging_Derivative}
Taking derivatives is often easier than integrating. The same holds for average integrals of the dynamical map, where we rephrase the average integral in terms of a differential operator defined by the characteristic function $\phi(k)$ of the underlying distribution. Despite differentiating non-commuting matrix exponentials being nontrivial, it offers a systematic insight into how averaging influences the dynamics.
\begin{align}
\notag
    \Lambda_{t}&=\int\ d^{n}\lambda\ p(\lambda)\mathcal{U}_{\lambda}(t)\\\notag
    &=\int\ d^{n}\lambda\ \frac{d^{n}k}{(2\pi)^{\frac{n}{2}}}\ \phi(k)e^{-ik\cdot\lambda} \mathcal{U}_{\lambda}(t)\\\notag
     &=\int\ d^{n}\lambda\ \frac{d^{n}k}{(2\pi)^{\frac{n}{2}}}\ \phi(k)e^{-ik\cdot\lambda} \left.e^{ik\cdot\mu}\right|_{\mu=0}\mathcal{U}_{\lambda}(t)\\\notag
    &=\int\ d^{n}\lambda\ \ \frac{d^{n}k}{(2\pi)^{\frac{n}{2}}}\ \left.\phi(-i\nabla_{\mu})\right|_{\mu=0}e^{ik\cdot(\mu-\lambda)} \ \mathcal{U}_{\lambda}(t)\\\notag
    &=\left.\phi(-i\nabla_{\mu})\right|_{\mu=0}\int\ d^{n}\lambda\ \delta(\mu-\lambda)\mathcal{U}_{\lambda}(t)\\\notag
    &=\left.\phi(-i\nabla_{\mu})\ \mathcal{U}_{\mu}(t)\right|_{\mu=0}\\
    &\equiv D\mathcal{U}(t).
\end{align}
In general, any quantity $f(\lambda)$ can be averaged by $\bar{f}=Df$.

\subsection{Example: Solving the SK model}
\label{app:exampleSKModel}
First, we consider an integrable model for illustration purposes, which is also solvable by other means than the differential one. A well-known example is the SK model, an Ising model with independently Gaussian distributed all-to-all couplings $H_{\lambda}=\sum_{i<j}\lambda_{ij}Z_{i}Z_{j}$, $\lambda_{ij}\in \mathcal{N}(J_{ij},\sigma_{ij}^2)\ \forall i,j$.
The Hamiltonian is a sum of commuting operators. Thus, also any pair of commutators of these operators commute, i.e.\ $\forall i,j,k,l:\ [[Z_{i}Z_{j},\ \cdot\ ],[Z_{k}Z_{l},\ \cdot\ ]]=0$. Calculating the dynamical map is now straightforward. The characteristic function is $\phi(k_{12},\dots,k_{(N-1)N})=\prod_{i<j}\exp{\left\{-\sigma_{ij}^{2}k_{ij}^2/2+iJ_{ij}k_{ij}\right\}}$ and therefore 
\begin{align}
\notag
\Lambda_t&=\left.\prod_{i<j}e^{\frac{\sigma_{ij}^{2}}{2}\partial_{\mu_{ij}}^2+J_{ij}\partial_{\mu_{ij}}}\ \mathcal{U}_{\mu}(t)\right|_{\mu=0}\\\notag
&=\left.\prod_{i<j}e^{\frac{\sigma_{ij}^{2}}{2}\partial_{\mu_{ij}}^2+J_{ij}\partial_{\mu_{ij}}}e^{-it\mu_{ij}[Z_{i}Z_{j},\ \cdot\ ]}\right|_{\mu=0}\\\notag
&=\prod_{i<j}e^{-\frac{\sigma_{ij}^{2}t^{2}}{2}[Z_{i}Z_{j},\ \cdot\ ]^{2}-itJ_{ij}[Z_{i}Z_{j},\ \cdot\ ]}\\
&=\exp{\left\{\sum_{i<j}-\frac{\sigma_{ij}^{2}t^{2}}{2}[Z_{i}Z_{j},\ \cdot\ ]^{2}\right\}}e^{-it[\bar{H},\ \cdot\ ]}.
\end{align}
We can easily see that the unitary dynamics is governed by the average Hamiltonian, while decoherence is caused by the first exponential. A Lindbladian form can easily be read off:
\begin{align}
\notag
\mathcal{L}_{t}=&-i[\bar{H},\ \cdot\ ]\\
&+\sum_{i<j}2\sigma_{ij}^{2}t\left[Z_{i}Z_{j}\cdot Z_{i}Z_{j} -\frac{1}{2}\left\{(Z_{i}Z_{j})^2,\ \cdot\ \right\}\right]
\end{align}
We note some interesting properties, which can be deduced from this Lindbladian:
\begin{enumerate}
    \item If all $\mu_{ij}=\mu$ and $\sigma_{ij}^2=\sigma^2$ such that the ensemble is permutation invariant, then $\Lambda_t$ and $\mathcal{L}_t$ are composed of only the permutation invariant operators $\sum_{i<j}[Z_{i}Z_{j},\ \cdot\ ]$ and $\sum_{i<j} [Z_{i}Z_{j},\ \cdot\ ]^2$. In this way, the symmetry of the ensemble directly manifests itself in the structure of the effective time evolution.
    \item Noise in the coupling parameters $\lambda_{ij}$  leads to two-body decoherence terms.
    \item Gaussian noise always leads to time-dependent, linearly increasing decoherence rates.
\end{enumerate}

\subsection{Weak-disorder expansion}
\label{app:derivation_weak_time_evolution}
Application of the operator $D$ to $\mathcal{U}(t)$ becomes non-trivial as soon as non-commuting operators are involved. To benchmark our new tool, we add a constant transverse field to the SK model. Splitting $D$ into its coherent and incoherent part, i.e. splitting off the complex phase of the characteristic function, where $\Sigma=\mathrm{diag}(\sigma^2_{12},\dots,\sigma^2_{n-1,n})$ yields 
\begin{align}
\label{eq:WDE_translation}
\notag
    \Lambda_{t}&=\left.\exp{\left\{\frac{1}{2}\nabla_{\mu}^{T}\cdot\Sigma\cdot \nabla_{\mu}\right\}e^{J \cdot \nabla_{\mu}}}\mathcal{U}_{\mu}(t)\right|_{\mu=0}\\\notag
    &=\left.\exp{\left\{\frac{1}{2}\nabla_{\mu}^{T}\cdot\Sigma\cdot \nabla_{\mu}\right\}}\mathcal{U}_{\mu+J}(t)\right|_{\mu=0}\\\notag
    &=\left.\exp{\left\{\frac{1}{2}\nabla_{\mu+J}^{T}\cdot\Sigma\cdot \nabla_{\mu+J}\right\}}\mathcal{U}_{\mu+J}(t)\right|_{\mu=0}\\\notag
    &=\left.\exp{\left\{\frac{1}{2}\nabla_{\mu}^{T}\cdot\Sigma\cdot \nabla_{\mu}\right\}}\mathcal{U}_{\mu}(t)\right|_{\mu=J}\\\notag
    &=\left.\exp{\left\{\sum_{i<j}\frac{\sigma_{ij}^2}{2}\partial_{\mu_{ij}}^{2}\right\}}\mathcal{U}_{\mu}(t)\right|_{\mu=J}\\
    &\equiv \tilde{D}\mathcal{U}(t).
\end{align}
To make further progress, a discussion on differentiation of non-commuting matrix exponentials is in order. The first partial derivative is given by
\begin{align}
\notag
\partial_{\mu_{ij}}e^{-it[H_{\mu},\ \cdot\ ]}&=-it\int_{0}^{1}\ &&\hspace{-0.4cm}dx\ e^{-it(1-x)[H_{\mu},\ \cdot\ ]} (\partial_{\mu_{ij}}[H_{\mu},\ \cdot\ ])\\\notag&
&&\hspace{-0.4cm}\times e^{-itx[H_{\mu},\ \cdot\ ]}\\
&=-i\ &&\hspace{-1.2cm}\int_{0}^{t}\ dt'\ \mathcal{U}_{\mu}(t-t')[Z_{i}Z_{j},\ \cdot\ ]\mathcal{U}_{\mu}(t').
\end{align}
The action of higher-order partial derivatives can already be read off because another differentiation acts either on the first or the second time evolution operator leading to a second integral, while the Leibniz rule causes a sum over both cases resulting in a time ordering of the operators that correspond to the indices of the partial derivative:
\begin{align}
\notag
&\phantom{=}\partial_{\mu_{kl}}\partial_{\mu_{ij}}e^{-it[H_{\mu},\ \cdot\ ]}=\\\notag
&\phantom{=}(-i)^2 \int_{0}^{t}\ dt_{1}\left\{\phantom{+}\int_{0}^{t-t_{1}}\hspace{-0.3cm}dt_{2}\ \mathcal{U}_{\mu}(t-t_{1}-t_{2})[Z_{k}Z_{l},\ \cdot\ ]
 \mathcal{U}_{\mu}(t_{2})\right.\\\notag&\hspace{3.4cm}\times[Z_{i}Z_{j},\ \cdot\ ]\mathcal{U}_{\mu}(t_{1})\\\notag
&\phantom{=}\hspace{2.4cm}+\int_{0}^{t_{1}}\hspace{-0.1cm}dt_{2}\ \mathcal{U}_{\mu}(t-t_{1})[Z_{i}Z_{j},\ \cdot\ ]\mathcal{U}_{\mu}(t_{1}-t_{2})\\\notag
&\phantom{=}\left.\hspace{3.4cm}\times[Z_{k}Z_{l},\ \cdot\ ] \mathcal{U}_{\mu}(t_{2})\right\}\\\notag
&=(-i)^2\int_{0}^{t}\hspace{-0.1cm} dt_{1}\ \int_{0}^{t_{1}}\hspace{-0.1cm}dt_{2} \left\{\phantom{+}\mathcal{U}_{\mu}(t-t_{1})[Z_{i}Z_{j},\ \cdot\ ]\mathcal{U}_{\mu}(t_{1})\right.\\\notag
&\phantom{=}\hspace{3.4cm}\phantom{+}\times\mathcal{U}_{\mu}(-t_{2})[Z_{k}Z_{l},\ \cdot\ ]\mathcal{U}_{\mu}(t_{2})\\\notag
&\phantom{=}\hspace{3.4cm}+\mathcal{U}_{\mu}(t-t_{1})[Z_{k}Z_{l},\ \cdot\ ]\mathcal{U}_{\mu}(t_{1})\\\notag
&\left.\phantom{=}\hspace{3.4cm}\phantom{+}\times\mathcal{U}_{\mu}(-t_{2})[Z_{i}Z_{j},\ \cdot\ ]\mathcal{U}_{\mu}(t_{2})\right\}\\\notag
&=(-i)^2\mathcal{U}_{\mu}(t)\int_{0}^{t}\hspace{-0.1cm}dt_{1}\hspace{-0.1cm}\int_{0}^{t_{1}}\hspace{-0.1cm}dt_{2}[\{Z_{i}Z_{j}\}_{t_{1}}^{\mu},\ \cdot\ ][\{Z_{k}Z_{l}\}_{t_{2}}^{\mu},\ \cdot\ ]\\
&\phantom{=}\hspace{3.4cm}+[\{Z_{k}Z_{l}\}_{t_{1}}^{\mu},\ \cdot\ ][\{Z_{i}Z_{j}\}_{t_{2}}^{\mu},\ \cdot\ ].
\end{align}
Note that $\mathcal{U}_{\mu}(t)$ will always appear at the leftmost operator and thus can always be pulled out to the left, resulting in a product of the time evolution governed by the average Hamiltonian with a time-ordered series of commutators, which are evolved with respect to the average Hamiltonian themselves, after setting $\mu=J$.

Returning to the symmetric case where all couplings experience the same distribution, we approximate $\Lambda_{t}$ for weak disorder by truncating the expansion of $\tilde{D}$ at order $\sigma^{2}$. The resulting formula is then given by Eq.~\eqref{eq:WDE_Lambda_t}. To calculate $[Z_{i}Z_{j}(t_{1}),\ \cdot\ ][Z_{i}Z_{j}(t_{2}),\ \cdot\ ]$ in the symmetric sector $\bar{\mathcal{U}}(t)$ needs to be known. Therefore, $\exp{\{-it\bar{H}\}}$ ought to be analytically solvable. As an example, we use the case $\mu=0$, then $\bar{\mathcal{U}}(t)=\exp{\{-ith\sum_{k}[X_{k},\ \cdot\ ]\}}$ and
\begin{align}
\notag
[Z_{i}Z_{j}(t),\ \cdot\ ]=&\phantom{+}\cos^{2}{(2ht)}[Z_{i}Z_{j},\ \cdot\ ]\\\notag
&+\frac{\sin{(4ht)}}{2}\left([Y_{i}Z_{j},\ \cdot\ ]+[Z_{i}Y_{j},\ \cdot\ ]\right)\\\notag
&+\sin^{2}{(2ht)}[Y_{i}Y_{j},\ \cdot\ ].
\end{align}
After taking the product and summing over $i<j$ the resulting operator becomes symmetric and can be reduced to the permutation-invariant subspace.

\section{PDE perspective}
\label{app:PDE_perspective}
More insight can be gained through the lens of partial differential equations. We may neglect the projection at the end of the averaging procedure and leave $\mu$ an arbitrary parameter, i.e.\ we do not set $\mu=0$ or $\mu=J$. Thus, we obtain an effective description in the tensor space of functions and Hermitian matrices:
\begin{align}
\label{app:eq.:average_derivative_general1}
&\bar\rho= \int \mathrm{d}p(\lambda) U_{\lambda}(t)\rho_0U_{\lambda}^\dagger(t) = |\varphi|(i\nabla_{\mu})\rho_{\mu}(t)|_{\mu=J}\\
\label{app:eq.:average_derivative_general2}
\Rightarrow &\bar{\rho}_{\mu}=|\varphi|(i\nabla_{\mu})\rho_{\mu}(t)\\\notag
\Rightarrow &\dot{\bar{\rho}}_{\mu}=\{-i[H_{\mu},\ \cdot\ ]-[\partial_{\mu_{ij}}H,\ \cdot\ ](\partial_{k_{ij}}\log{|\varphi|})(-i\nabla_{\mu})\}\bar{\rho}_{\mu}\\\notag
&\phantom{\dot{\bar{\rho}}_{\mu}}=-i[Z_{i}Z_{j},\ \cdot\ ](\mu_{ij}+\sigma^{2}\partial_{\mu_{ij}})\bar{\rho}_{\mu}\\\label{app:eq.:average_derivative_general3}
&\phantom{\dot{\bar{\rho}}_{\mu}}\equiv\mathcal{L}_{\mu}\bar{\rho}_{\mu}
\end{align}
Equations~\eqref{app:eq.:average_derivative_general1}~and~\eqref{app:eq.:average_derivative_general2} hold generally for all symmetric distributions, since the complex phase of their characteristic function will be $\exp(ikJ)$.
Viewed on the space $\mathds{R}[\mu]\otimes \mathcal{B}(\mathcal{H})$, time evolution has incoherent contributions only on the function space (analogous to harmonic oscillators we can understand $\mu_{ij}+\sigma^2\partial_{\mu_{ij}}=a_{ij}$ as mathematically similar to an annihilation operator in function space, and therefore this is the only non-Hermitian part in the effective time evolutions generator). Equation~\eqref{app:eq.:average_derivative_general3} implies non-separability between the operator space and polynomial space of the auxiliary variables $\mu_{ij}$, i.e. $\mathds{R}[\mu]$, for the resulting state $\bar{\rho}_{\mu}(t)$. This is because the symmetry of the distribution is fundamentally living within the tensor space $\mathrm{End}(\mathds{R}[\mu])\otimes\mathrm{End}(\mathcal{B}(\mathcal{H}))$. Setting the dummy variable to a fixed value, as is necessary to perform the average, the elements of $\mathrm{End}(\mathds{R}[\mu])\otimes\mathrm{End}(\mathcal{B}(\mathcal{H}))$ are effectively projected onto superoperators, i.e.\ it is a projection $\Pi:\mathrm{End}(\mathds{R}[\mu])\otimes\mathrm{End}(\mathcal{B}(\mathcal{H}))\rightarrow \mathrm{End}(\mathcal{B}(\mathcal{H}))$. To perform the projection, the action of $\partial_{\mu_{ij}}$ onto $\bar{\rho}_{\mu}$ is needed explicitly and has to be approximated.

\section{Symmetric Pauli Strings and Projection Combinatorics}
\label{app:symmetric_pauli_strings}
The set of all totally symmetric Pauli strings is a sub-vector space of the space of Hermitian matrices. For $N$ spins, there are $\binom{N+3}{3}$ linearly independent strings, which are labeled by $(x,y,z)$ and defined via 
\begin{align}
    \Sigma_{(x,y,z)}=\mathcal{N}_{xyz}^{-\frac{1}{2}}\sum_{s\in \tilde{S}_{N}}\prod_{i=1}^{x}X_{s(i)}\ \prod_{j=x+1}^{x+y}Y_{s(j)}\prod_{k=x+y+1}^{x+y+z}Z_{s(k)},
\end{align}
where we define $\tilde{S}_{N}$ as the subgroup of all true permutations, i.e.\ the symmetric group modulo the point stabilizer of a Pauli string with $x$ $X$, $y$ $\hat
Y$ and $z$ $Z$ operators. Then, $\mathcal{N}_{xyz}$ is defined with respect to the Frobenius norm
\begin{align}
    \mathcal{N}_{xyz}=2^{N}\frac{N!}{x!y!z!(N-x-y-z)!}=2^N\binom{N}{x,y,z}.
\end{align}
Working in this basis, the projection of operators is straightforward. We only need to think about how many sites an operator acts on and which new $(x',y',z')$ are obtained. An example is given in the next appendix.

\section{Short-Time Expansion Projection}
\label{app:Lindbladterms}
The first three orders of the Lindblad expansion for the SK model with transverse field are given by
\begin{align}
&\kappa_{1}=\frac{J}{\sqrt{N}} \sum_{i<j}[Z_{i}Z_{j},\ \cdot\ ]+h\sum_{k}[X_{k},\ \cdot\ ],\\
&\kappa_{2}=\frac{\sigma^{2}}{N}\sum_{i<j}[Z_{i}Z_{j},\ \cdot\ ]^{2},\\
&\kappa_{3}=\frac{\sigma^{2}}{N} h \sum_{i\neq j}[Z_{i}Z_{j},\ \cdot\ ]^{2}[X_{j},\ \cdot\ ]
\end{align}
according to Eq.~\eqref{eq.:Lindblad_cumulants}. Instead of the exact summation we may choose a representative and look at its action on the symmetric subspace, since after projection the explicit indices of the representative used become irrelevant. Thus, the sum only contributes by a combinatorial factor
\begin{align}
\notag
    \kappa&_{1}|_{(x',y',z')}^{(x,y,z)}\\\notag
    =&\Tr\left(\Sigma_{(x',y',z')}\left[\mu \binom{N}{2}Z_{1}Z_{2}+Nh X_{1},\Sigma_{(x,y,z)}\right]\right)\\\notag
    =&\phantom{+}2J\left(\hspace{0.55cm}\delta_{(x',y',z')}^{(x-1,y+1,z+1)}\sqrt{x (y+1)(z+1) (N-x-y-z)}\right.\\\notag
    &\qquad\quad-\delta_{(x',y',z')}^{(x+1,y-1,z+1)}\sqrt{(x+1)y(z+1)(N-x-y-z)}\\\notag
    &\qquad\quad+\delta_{(x',y',z')}^{(x-1,y+1,z-1)}\sqrt{x(y+1)z(N-x-y-z+1)}\\\notag
    &\qquad\quad-\left.\delta_{(x',y',z')}^{(x+1,y-1,z-1)}\sqrt{(x+1)yz(N-x-y-z+1)}\right)\\\notag
    &+2h\left(\hspace{0.42cm}\delta_{(x',y',z')}^{(x,y-1,z+1)}\sqrt{y(z+1)}\right.\\\notag
    &\qquad\quad-\left.\delta_{(x',y',z')}^{(x,y+1,z-1)}\sqrt{(y+1)z}\right).
\end{align}
Here, the Kronecker deltas ensure the right mapping between the symmetric string indices. Applying the same procedure to the other operators leads to 
\begin{align}
    \notag
    &\mathcal{L}(t)_{(x',y',z')}^{(x,y,z)}=&&\\\notag
    &\phantom{+}2J\left(\hspace{0.55cm}\delta_{(x',y',z')}^{(x-1,y+1,z+1)}\sqrt{x (y+1)(z+1) (N-x-y-z)}\right.\\\notag
    &\qquad\quad-\delta_{(x',y',z')}^{(x+1,y-1,z+1)}\sqrt{(x+1)y(z+1)(N-x-y-z)}\\\notag
    &\qquad\quad+\delta_{(x',y',z')}^{(x-1,y+1,z-1)}\sqrt{x(y+1)z(N-x-y-z+1)}\\\notag
    &\qquad\quad-\left.\delta_{(x',y',z')}^{(x+1,y-1,z-1)}\sqrt{(x+1)yz(N-x-y-z+1)}\right)\\\notag
    &+2h\left(\hspace{0.42cm}\delta_{(x',y',z')}^{(x,y-1,z+1)}\sqrt{y(z+1)}\right.\\\notag
    &\qquad\quad-\left.\delta_{(x',y',z')}^{(x,y+1,z-1)}\sqrt{(y+1)z}\right)\\\notag
    &-4\sigma^{2}t\delta_{(x',y',z')}^{(x,y,z)}(x+y)(N-x-y)\\\notag
    &+4\sigma^{2}ht^{2}\left(\hspace{0.435cm}\delta_{(x',y',z')}^{(x,y-1,z+1)}(x+y-1)\hspace{0.75cm}\sqrt{y(z+1)}\right.\\\notag
    &\qquad\qquad\quad \left.-\delta_{(x',y',z')}^{(x,y+1,z-1)}(N-x-y-1)\sqrt{(y+1)z}\right)\\
    &+\mathcal{O}(t^{3})
\end{align}
for the Lindbladian to second order in time. The third order was computed numerically in an analogous way and is not included here explicitly to avoid excessively long expressions. Note that since only a representative's action is required, we can calculate it for the smallest possible system size. For example, if we are interested in the action of an operator acting on $k$ sites, then numerically only the $k$-particle Hilbert space is needed for the projection. Afterwards, we can embed the operator in the full $N$-particle space, because the coupling structure in the string indices $x$,$y$, and $z$ is independent of $N$, while the linear coefficients are simply adapted by division and multiplication of the corresponding combinatorial factors.

\section{Supplementary results on the weak-disorder expansion} 
\label{app:X_magnetization}
\begin{figure}
    \centering
    \includegraphics[width=0.4\textwidth]{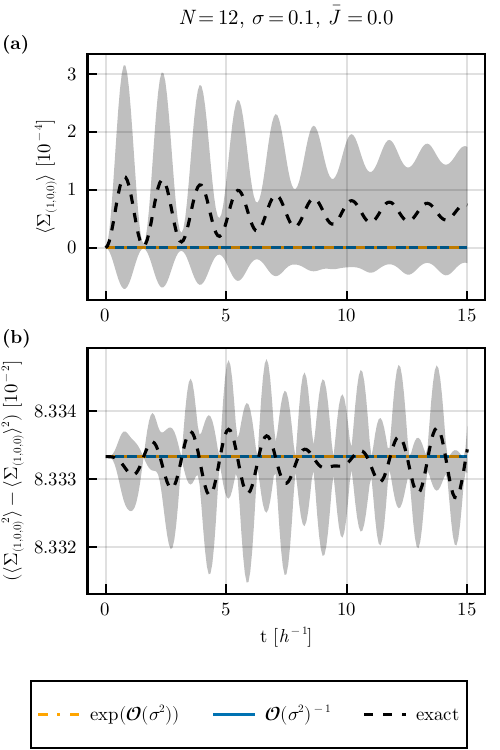}
    \caption{Weak-disorder expansion of transverse field Ising model. Exponential (orange) and inverse (blue) regularization schemes are plotted together with exact calculations (black) for $N=12$ spins. Panel (a) shows the total $X$ magnetization and (b) the corresponding variances for a $Z$-polarized initial state calculated for $10000$ shots. }\label{fig:x-magnetization}
\end{figure}

Figure~\ref{fig:x-magnetization} shows the $X$-magnetization and its variance for an initially $Z$-polarized state approximated by the weak-disorder expansion (truncated after the first order in $\sigma^2$) together with Monte-Carlo averaged results obtained with exact diagonalization. The plot illustrates the potential of the perturbative treatment with exact disorder averaging. For the TFIM model we can analytically show that the magnetization is constant. This is already captured accordingly at first order in $\sigma^2$. Monte-Carlo averaging, however, has problems to approach constant values due to statistical errors, even for a high sample number.
Moreover, the plot shows that the quality of the approximation at a given approximation order strongly depends on the initial state and the observable. In contrast to Fig.~\ref{fig:weak_disorder_expansionN12}, where the predicted $Z$-magnetization deviates at late times, the time evolution is correctly captured at all times.

\begin{figure}
    \centering
    \includegraphics[width=0.49\textwidth]{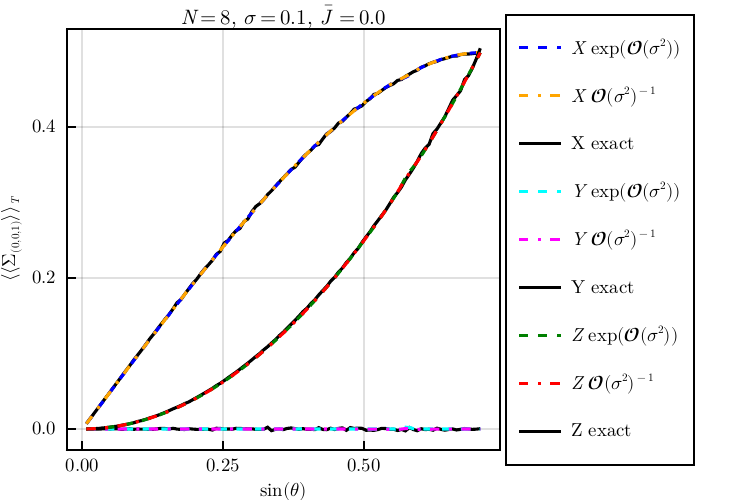}
    \caption{Weak-disorder expansion of transverse field Ising model with tilted field. Exponential and inverse regularization schemes are plotted together with exact calculations (black) for $N=8$ spins. The graph shows the average late-time $Z$ magnetization plotted against $\sin(\theta)$ for a $X$-, $Y$- and $Z$-polarized initial state calculated for $100$ shots.}
    \label{fig:late-time}
\end{figure}

To study late-time behavior, a tilted field is used such that the time-averaged expectation value may change for fixed initial polarization and different angles:
\begin{equation*}
    H=\sum_{i<j}\frac{J_{ij}}{\sqrt{N}}Z_{i}Z_{j}+h\sum_{i} \left(\cos{\theta}\ X_{i}+ \sin{\theta}\ Z_{i}\right).
\end{equation*}
We define the time-averaged expectation value 
\begin{align}
    \langle\langle\Sigma_{(0,0,1)}\rangle\rangle_{T}:=\frac{2}{T}\int_{\frac{T}{2}}^T dt' \langle\Sigma_{(0,0,1)}\rangle(t')
\end{align}
and plot it against $\sin(\theta)$ for the three cases of $X$-, $Y$- and $Z$-polarized initial state.
In Fig.~\ref{fig:late-time} we show the result for $N=8$ particles and $T=500h^{-1}$ using the exponential and the inverse regularization for weak disorder of Fig.~\ref{fig:weak_disorder_expansionN12} in Sec.~\ref{sec:weak_disorder_expansion} together with exact solutions for $100$ disorder shots. The weak-disorder approximations stay close to the late-time-averaged expectation value for all angles. Visible deviations are the result of sampling noise. The result was also confirmed for $N=3$ particles.

\newpage
\bibliography{main}

@article{Ritort01062003,
author = {F. Ritort and P. Sollich},
title = {Glassy dynamics of kinetically constrained models},
journal = {Advances in Physics},
volume = {52},
number = {4},
pages = {219--342},
year = {2003},
publisher = {Taylor \& Francis},
doi = {10.1080/0001873031000093582}
}

@article{PhysRevLett.53.958,
  title = {Models of Hierarchically Constrained Dynamics for Glassy Relaxation},
  author = {Palmer, R. G. and Stein, D. L. and Abrahams, E. and Anderson, P. W.},
  journal = {Phys. Rev. Lett.},
  volume = {53},
  issue = {10},
  pages = {958--961},
  numpages = {0},
  year = {1984},
  month = {Sep},
  publisher = {American Physical Society},
  doi = {10.1103/PhysRevLett.53.958},
  url = {https://link.aps.org/doi/10.1103/PhysRevLett.53.958}
}

@article{PhysRevLett.53.1244,
  title = {Kinetic Ising Model of the Glass Transition},
  author = {Fredrickson, Glenn H. and Andersen, Hans C.},
  journal = {Phys. Rev. Lett.},
  volume = {53},
  issue = {13},
  pages = {1244--1247},
  numpages = {0},
  year = {1984},
  month = {Sep},
  publisher = {American Physical Society},
  doi = {10.1103/PhysRevLett.53.1244},
  url = {https://link.aps.org/doi/10.1103/PhysRevLett.53.1244}
}

@article{Lee2025,
  doi = {10.22331/q-2025-01-23-1607},
  url = {https://doi.org/10.22331/q-2025-01-23-1607},
  title = {Symmetry protected topological phases under decoherence},
  author = {Lee, Jong Yeon and You, Yi-Zhuang and Xu, Cenke},
  journal = {{Quantum}},
  issn = {2521-327X},
  publisher = {{Verein zur F{\"{o}}rderung des Open Access Publizierens in den Quantenwissenschaften}},
  volume = {9},
  pages = {1607},
  month = jan,
  year = {2025}
}

@Article{Sarkar1987,
  Title                    = {Optical Bistability with Small Numbers of Atoms},
  Author                   = {S. Sarkar and J. S. Satchell},
  Journal                  = {EPL},
  Year                     = {1987},
  Number                   = {7},
  Pages                    = {797},
  Volume                   = {3},
  Url                      = {http://stacks.iop.org/0295-5075/3/i=7/a=005}
}

@Article{Hartmann2012,
  Title                    = {Generalized {D}icke States},
  Author                   = {S. Hartmann},
  journal = {Quantum Inf. Comput.},
  volume = {16},
  issue = {15},
  pages = {1333--1348},
  numpages = {16},
  year = {2016},
url = {https://dl.acm.org/doi/10.5555/3179439.3179444}
}

@article{Xu2013,
  title = {Simulating open quantum systems by applying {SU}(4) to quantum master equations},
  author = {Xu, Minghui and Tieri, D. A. and Holland, M. J.},
  journal = {Phys. Rev. A},
  volume = {87},
  issue = {6},
  pages = {062101},
  numpages = {7},
  year = {2013},
  month = {Jun},
  publisher = {American Physical Society},
  doi = {10.1103/PhysRevA.87.062101},
  url = {http://link.aps.org/doi/10.1103/PhysRevA.87.062101}
}

@article{Lipkin1965,
	title = {Validity of many-body approximation methods for a solvable model: (I). Exact solutions and perturbation theory},
	volume = {62},
	issn = {0029-5582},
	url = {https://www.sciencedirect.com/science/article/pii/002955826590862X},
	doi = {https://doi.org/10.1016/0029-5582(65)90862-X},
	pages = {188--198},
	number = {2},
	journal = {Nuclear Physics},
	author = {Lipkin, H. J. and Meshkov, N. and Glick, A. J.},
	year = {1965},
}

@article{Zhang2017,
	title = {Observation of a many-body dynamical phase transition with a 53-qubit quantum simulator},
	volume = {551},
	issn = {1476-4687},
	url = {https://doi.org/10.1038/nature24654},
	doi = {10.1038/nature24654},
	pages = {601--604},
	number = {7682},
	journal = {Nature},
	shortjournal = {Nature},
	author = {Zhang, J. and Pagano, G. and Hess, P. W. and Kyprianidis, A. and Becker, P. and Kaplan, H. and Gorshkov, A. V. and Gong, Z.-X. and Monroe, C.},
	year = {2017},
}

@article{Muniz2020,
	title = {Exploring dynamical phase transitions with cold atoms in an optical  cavity},
	volume = {580},
	issn = {1476-4687},
	url = {https://doi.org/10.1038/s41586-020-2224-x},
	doi = {10.1038/s41586-020-2224-x},
	pages = {602--607},
	number = {7805},
	journal = {Nature},
	shortjournal = {Nature},
	author = {Muniz, Juan A. and Barberena, Diego and Lewis-Swan, Robert J. and Young, Dylan J. and Cline, Julia R. K. and Rey, Ana Maria and Thompson, James K.},
	year = {2020},
}

@article{Dusuel2004,
  title = {Finite-Size Scaling Exponents of the Lipkin-Meshkov-Glick Model},
  author = {Dusuel, S\'ebastien and Vidal, Julien},
  journal = {Phys. Rev. Lett.},
  volume = {93},
  issue = {23},
  pages = {237204},
  numpages = {4},
  year = {2004},
  month = {Dec},
  publisher = {American Physical Society},
  doi = {10.1103/PhysRevLett.93.237204},
  url = {https://link.aps.org/doi/10.1103/PhysRevLett.93.237204}
}

@article{Latorre2005,
  title = {Entanglement entropy in the Lipkin-Meshkov-Glick model},
  author = {Latorre, Jos\'e I. and Or\'us, Rom\'an and Rico, Enrique and Vidal, Julien},
  journal = {Phys. Rev. A},
  volume = {71},
  issue = {6},
  pages = {064101},
  numpages = {4},
  year = {2005},
  month = {Jun},
  publisher = {American Physical Society},
  doi = {10.1103/PhysRevA.71.064101},
  url = {https://link.aps.org/doi/10.1103/PhysRevA.71.064101}
}

@article{Ma2011,
	title = {Quantum spin squeezing},
	volume = {509},
	issn = {0370-1573},
	url = {https://www.sciencedirect.com/science/article/pii/S0370157311002201},
	doi = {https://doi.org/10.1016/j.physrep.2011.08.003},
	pages = {89--165},
	number = {2},
	journal = {Physics Reports},
	author = {Ma, Jian and Wang, Xiaoguang and Sun, C. P. and Nori, Franco},
	year = {2011},
}

@article{Hepp1973,
  title = {Equilibrium Statistical Mechanics of Matter Interacting with the Quantized Radiation Field},
  author = {Hepp, Klaus and Lieb, Elliott H.},
  journal = {Phys. Rev. A},
  volume = {8},
  issue = {5},
  pages = {2517--2525},
  numpages = {0},
  year = {1973},
  month = {Nov},
  publisher = {American Physical Society},
  doi = {10.1103/PhysRevA.8.2517},
  url = {https://link.aps.org/doi/10.1103/PhysRevA.8.2517}
}

@article{Dicke1954,
  title = {Coherence in Spontaneous Radiation Processes},
  author = {Dicke, R. H.},
  journal = {Phys. Rev.},
  volume = {93},
  issue = {1},
  pages = {99--110},
  numpages = {0},
  year = {1954},
  month = {Jan},
  publisher = {American Physical Society},
  doi = {10.1103/PhysRev.93.99},
  url = {https://link.aps.org/doi/10.1103/PhysRev.93.99}
}

@article{Maldacena2016,
  title = {Remarks on the Sachdev-Ye-Kitaev model},
  author = {Maldacena, Juan and Stanford, Douglas},
  journal = {Phys. Rev. D},
  volume = {94},
  issue = {10},
  pages = {106002},
  numpages = {43},
  year = {2016},
  month = {Nov},
  publisher = {American Physical Society},
  doi = {10.1103/PhysRevD.94.106002},
  url = {https://link.aps.org/doi/10.1103/PhysRevD.94.106002}
}

@MISC{Kitaev2015,
       author = {{Kitaev}, Alexei},
        title = "{A simple model of quantum holography (Apr 7 2015))}",
 howpublished = {Talks given at The Kavli Institute for Theoretical Physics Program: Entanglement in Strongly-Correlated Quantum Matter (Apr 6 - Jul 2, 2015)},
         year = 2015,
        month = apr,
          eid = {2},
        pages = {2}
}

@article{Sachdev1993,
  title = {Gapless spin-fluid ground state in a random quantum Heisenberg magnet},
  author = {Sachdev, Subir and Ye, Jinwu},
  journal = {Phys. Rev. Lett.},
  volume = {70},
  issue = {21},
  pages = {3339--3342},
  numpages = {0},
  year = {1993},
  month = {May},
  publisher = {American Physical Society},
  doi = {10.1103/PhysRevLett.70.3339},
  url = {https://link.aps.org/doi/10.1103/PhysRevLett.70.3339}
}

@article{Abanin2019,
  title = {Colloquium: Many-body localization, thermalization, and entanglement},
  author = {Abanin, Dmitry A. and Altman, Ehud and Bloch, Immanuel and Serbyn, Maksym},
  journal = {Rev. Mod. Phys.},
  volume = {91},
  issue = {2},
  pages = {021001},
  numpages = {26},
  year = {2019},
  month = {May},
  publisher = {American Physical Society},
  doi = {10.1103/RevModPhys.91.021001},
  url = {https://link.aps.org/doi/10.1103/RevModPhys.91.021001}
}

@article{Antinucci2023, 
title={{Symmetries and topological operators, on average}}, 
author={Andrea Antinucci and Giovanni Galati and Giovanni Rizi and Marco Serone}, 
journal={SciPost Phys.}, 
volume={15}, 
pages={125}, 
year={2023}, 
publisher={SciPost}, 
doi={10.21468/SciPostPhys.15.3.125}, 
url={https://scipost.org/10.21468/SciPostPhys.15.3.125}, 
}

@article{Ma2023,
  title = {Average Symmetry-Protected Topological Phases},
  author = {Ma, Ruochen and Wang, Chong},
  journal = {Phys. Rev. X},
  volume = {13},
  issue = {3},
  pages = {031016},
  numpages = {24},
  year = {2023},
  month = {Aug},
  publisher = {American Physical Society},
  doi = {10.1103/PhysRevX.13.031016},
  url = {https://link.aps.org/doi/10.1103/PhysRevX.13.031016}
}

@article{deGroot2022,
  doi = {10.22331/q-2022-11-10-856},
  url = {https://doi.org/10.22331/q-2022-11-10-856},
  title = {Symmetry {P}rotected {T}opological {O}rder in {O}pen {Q}uantum {S}ystems},
  author = {de Groot, Caroline and Turzillo, Alex and Schuch, Norbert},
  journal = {{Quantum}},
  issn = {2521-327X},
  publisher = {{Verein zur F{\"{o}}rderung des Open Access Publizierens in den Quantenwissenschaften}},
  volume = {6},
  pages = {856},
  month = nov,
  year = {2022}
}

@article{Fulga2014,
  title = {Statistical topological insulators},
  author = {Fulga, I. C. and van Heck, B. and Edge, J. M. and Akhmerov, A. R.},
  journal = {Phys. Rev. B},
  volume = {89},
  issue = {15},
  pages = {155424},
  numpages = {6},
  year = {2014},
  month = {Apr},
  publisher = {American Physical Society},
  doi = {10.1103/PhysRevB.89.155424},
  url = {https://link.aps.org/doi/10.1103/PhysRevB.89.155424}
}

@article{Mong2012,
  title = {Quantum Transport and Two-Parameter Scaling at the Surface of a Weak Topological Insulator},
  author = {Mong, Roger S. K. and Bardarson, Jens H. and Moore, Joel E.},
  journal = {Phys. Rev. Lett.},
  volume = {108},
  issue = {7},
  pages = {076804},
  numpages = {5},
  year = {2012},
  month = {Feb},
  publisher = {American Physical Society},
  doi = {10.1103/PhysRevLett.108.076804},
  url = {https://link.aps.org/doi/10.1103/PhysRevLett.108.076804}
}

@article{Ringel2012,
  title = {Strong side of weak topological insulators},
  author = {Ringel, Zohar and Kraus, Yaacov E. and Stern, Ady},
  journal = {Phys. Rev. B},
  volume = {86},
  issue = {4},
  pages = {045102},
  numpages = {11},
  year = {2012},
  month = {Jul},
  publisher = {American Physical Society},
  doi = {10.1103/PhysRevB.86.045102},
  url = {https://link.aps.org/doi/10.1103/PhysRevB.86.045102}
}

@article{Ma2025,
  title = {Topological Phases with Average Symmetries: The Decohered, the Disordered, and the Intrinsic},
  author = {Ma, Ruochen and Zhang, Jian-Hao and Bi, Zhen and Cheng, Meng and Wang, Chong},
  journal = {Phys. Rev. X},
  volume = {15},
  issue = {2},
  pages = {021062},
  numpages = {33},
  year = {2025},
  month = {May},
  publisher = {American Physical Society},
  doi = {10.1103/PhysRevX.15.021062},
  url = {https://link.aps.org/doi/10.1103/PhysRevX.15.021062}
}

@article{Chinnarasu2025,
  title = {Variational Simulation of the Lipkin-Meshkov-Glick Model on a Neutral Atom Quantum Computer},
  author = {Chinnarasu, R. and Poole, C. and Phuttitarn, L. and Noori, A. and Graham, T. M. and Coppersmith, S. N. and Balantekin, A. B. and Saffman, M.},
  journal = {PRX Quantum},
  volume = {6},
  issue = {2},
  pages = {020350},
  numpages = {18},
  year = {2025},
  month = {Jun},
  publisher = {American Physical Society},
  doi = {10.1103/PRXQuantum.6.020350},
  url = {https://link.aps.org/doi/10.1103/PRXQuantum.6.020350}
}

@misc{mansky2023permutationinvariantquantumcircuits,
      title={Permutation-invariant quantum circuits}, 
      author={Maximilian Balthasar Mansky and Santiago Londoño Castillo and Victor Ramos Puigvert and Claudia Linnhoff-Popien},
      year={2023},
      eprint={2312.14909},
      archivePrefix={arXiv},
      primaryClass={quant-ph},
      url={https://arxiv.org/abs/2312.14909}, 
}

@inproceedings{Sandvik_2010,
   title={Computational Studies of Quantum Spin Systems},
   ISSN={0094-243X},
   url={http://dx.doi.org/10.1063/1.3518900},
   DOI={10.1063/1.3518900},
   booktitle={AIP Conference Proceedings},
   publisher={AIP},
   author={Sandvik, Anders W. and Avella, Adolfo and Mancini, Ferdinando},
   year={2010} }

@article{andersonabsenceofdiffusion,
  title = {Absence of Diffusion in Certain Random Lattices},
  author = {Anderson, P. W.},
  journal = {Phys. Rev.},
  volume = {109},
  issue = {5},
  pages = {1492--1505},
  numpages = {0},
  year = {1958},
  month = {Mar},
  publisher = {American Physical Society},
  doi = {10.1103/PhysRev.109.1492},
  url = {https://link.aps.org/doi/10.1103/PhysRev.109.1492}
}

@article{RevModPhys.58.801,
  title = {Spin glasses: Experimental facts, theoretical concepts, and open questions},
  author = {Binder, K. and Young, A. P.},
  journal = {Rev. Mod. Phys.},
  volume = {58},
  issue = {4},
  pages = {801--976},
  numpages = {0},
  year = {1986},
  month = {Oct},
  publisher = {American Physical Society},
  doi = {10.1103/RevModPhys.58.801},
  url = {https://link.aps.org/doi/10.1103/RevModPhys.58.801}
}

@misc{gestssonEquivalenceDynamicsDisordered2024,
  title = {Equivalence of Dynamics of Disordered Quantum Ensembles and Semi-Infinite Lattices},
  author = {Gestsson, Hallmann {\'O}skar and Nation, Charlie and {Olaya-Castro}, Alexandra},
  year = {2024},
  month = jun,
  number = {arXiv:2406.17865},
  eprint = {2406.17865},
  primaryclass = {cond-mat, physics:quant-ph},
  publisher = {arXiv},
  urldate = {2024-07-07},
  archiveprefix = {arXiv},
  langid = {english}
}

@article{gneitingDisorderdressedQuantumEvolution2020,
  title = {Disorder-Dressed Quantum Evolution},
  author = {Gneiting, Clemens},
  year = {2020},
  month = jun,
  journal = {Phys. Rev. B},
  volume = {101},
  number = {21},
  pages = {214203},
  issn = {2469-9950, 2469-9969},
  doi = {10.1103/PhysRevB.101.214203},
  urldate = {2022-07-18},
  langid = {english}
}

@article{gneitingDisorderRobustEntanglementTransport2019,
  title = {Disorder-{{Robust Entanglement Transport}}},
  author = {Gneiting, Clemens and Leykam, Daniel and Nori, Franco},
  year = {2019},
  month = feb,
  journal = {Phys. Rev. Lett.},
  volume = {122},
  number = {6},
  pages = {066601},
  issn = {0031-9007, 1079-7114},
  doi = {10.1103/PhysRevLett.122.066601},
  urldate = {2022-07-18},
  langid = {english}
}

@article{gneitingIncoherentEnsembleDynamics2016,
  title = {Incoherent Ensemble Dynamics in Disordered Systems},
  author = {Gneiting, Clemens and Anger, Felix R. and Buchleitner, Andreas},
  year = {2016},
  month = mar,
  journal = {Phys. Rev. A},
  volume = {93},
  number = {3},
  pages = {032139},
  issn = {2469-9926, 2469-9934},
  doi = {10.1103/PhysRevA.93.032139},
  urldate = {2022-07-18},
  langid = {english}
}

@article{gneitingQuantumEvolutionDisordered2017,
  title = {Quantum Evolution in Disordered Transport},
  author = {Gneiting, Clemens and Nori, Franco},
  year = {2017},
  month = aug,
  journal = {Phys. Rev. A},
  volume = {96},
  number = {2},
  pages = {022135},
  issn = {2469-9926, 2469-9934},
  doi = {10.1103/PhysRevA.96.022135},
  urldate = {2022-07-18},
  langid = {english}
}

@article{kropfEffectiveDynamicsDisordered2016,
  title = {Effective {{Dynamics}} of {{Disordered Quantum Systems}}},
  author = {Kropf, Chahan M. and Gneiting, Clemens and Buchleitner, Andreas},
  year = {2016},
  month = aug,
  journal = {Phys. Rev. X},
  volume = {6},
  number = {3},
  pages = {031023},
  issn = {2160-3308},
  doi = {10.1103/PhysRevX.6.031023},
  urldate = {2022-01-31},
  langid = {english}
}

@article{kropfOpenSystemModel2017,
  title = {Open System Model for Quantum Dynamical Maps with Classical Noise and Corresponding Master Equations},
  author = {Kropf, Chahan M. and Shatokhin, Vyacheslav N. and Buchleitner, Andreas},
  year = {2017},
  month = dec,
  journal = {Open Systems \& Information Dynamics},
  volume = {24},
  number = {04},
  eprint = {1711.01578},
  primaryclass = {cond-mat, physics:quant-ph},
  pages = {1740012},
  issn = {1230-1612, 1793-7191},
  doi = {10.1142/S1230161217400121},
  urldate = {2024-07-18},
  archiveprefix = {arXiv},
  langid = {english}
}

@article{paredesExploitingQuantumParallelism2005,
  title = {Exploiting {{Quantum Parallelism}} to {{Simulate Quantum Random Many-Body Systems}}},
  author = {Paredes, B. and Verstraete, F. and Cirac, J. I.},
  year = {2005},
  month = sep,
  journal = {Phys. Rev. Lett.},
  volume = {95},
  number = {14},
  pages = {140501},
  issn = {0031-9007, 1079-7114},
  doi = {10.1103/PhysRevLett.95.140501},
  urldate = {2024-08-08},
  copyright = {http://link.aps.org/licenses/aps-default-license},
  langid = {english}
}

@article{sherringtonSolvableModelSpinGlass1975,
  title = {Solvable {{Model}} of a {{Spin-Glass}}},
  author = {Sherrington, David and Kirkpatrick, Scott},
  year = {1975},
  month = dec,
  journal = {Phys. Rev. Lett.},
  volume = {35},
  number = {26},
  pages = {1792--1796},
  issn = {0031-9007},
  doi = {10.1103/PhysRevLett.35.1792},
  urldate = {2024-07-12},
  copyright = {http://link.aps.org/licenses/aps-default-license},
  langid = {english}
}

@article{kucskoCriticalThermalizationDisordered2018,
  title = {Critical {{Thermalization}} of a {{Disordered Dipolar Spin System}} in {{Diamond}}},
  author = {Kucsko, G. and Choi, S. and Choi, J. and Maurer, P. C. and Zhou, H. and Landig, R. and Sumiya, H. and Onoda, S. and Isoya, J. and Jelezko, F. and Demler, E. and Yao, N. Y. and Lukin, M. D.},
  year = {2018},
  month = jul,
  journal = {Phys. Rev. Lett.},
  volume = {121},
  number = {2},
  pages = {023601},
  issn = {0031-9007, 1079-7114},
  doi = {10.1103/PhysRevLett.121.023601},
  urldate = {2022-07-18},
  langid = {english}
}

@Article{Yan2013,
author={Yan, Bo
and Moses, Steven A.
and Gadway, Bryce
and Covey, Jacob P.
and Hazzard, Kaden R. A.
and Rey, Ana Maria
and Jin, Deborah S.
and Ye, Jun},
title={Observation of dipolar spin-exchange interactions with lattice-confined polar molecules},
journal={Nature},
year={2013},
month={Sep},
day={01},
volume={501},
number={7468},
pages={521-525},
issn={1476-4687},
doi={10.1038/nature12483},
url={https://doi.org/10.1038/nature12483}
}

@article{signolesGlassyDynamicsDisordered2021,
  title = {Glassy Dynamics in a Disordered {{Heisenberg}} Quantum Spin System},
  author = {Signoles, A. and Franz, T. and Alves, R. Ferracini and G{\"a}rttner, M. and Whitlock, S. and Z{\"u}rn, G. and Weidem{\"u}ller, M.},
  year = {2021},
  month = jan,
  journal = {Phys. Rev. X},
  volume = {11},
  number = {1},
  eprint = {1909.11959},
  pages = {011011},
  issn = {2160-3308},
  doi = {10.1103/PhysRevX.11.011011},
  urldate = {2021-11-23},
  archiveprefix = {arXiv},
  langid = {english}
}

@article{datserisDrWatsonPerfectSidekick2020,
  title = {{{DrWatson}}: The Perfect Sidekick for Your Scientific Inquiries},
  shorttitle = {{{DrWatson}}},
  author = {Datseris, George and Isensee, Jonas and Pech, Sebastian and G{\'a}l, Tam{\'a}s},
  year = {2020},
  month = oct,
  journal = {Journal of Open Source Software},
  volume = {5},
  number = {54},
  pages = {2673},
  issn = {2475-9066},
  doi = {10.21105/joss.02673},
  urldate = {2024-05-24},
  copyright = {http://creativecommons.org/licenses/by/4.0/}
}

@article{bezansonJuliaFreshApproach2017,
  title = {Julia: {{A Fresh Approach}} to {{Numerical Computing}}},
  shorttitle = {Julia},
  author = {Bezanson, Jeff and Edelman, Alan and Karpinski, Stefan and Shah, Viral B.},
  year = {2017},
  month = jan,
  journal = {SIAM Review},
  volume = {59},
  number = {1},
  pages = {65--98},
  issn = {0036-1445, 1095-7200},
  doi = {10.1137/141000671},
  urldate = {2022-07-22},
  langid = {english}
}

@misc{fonsvanderplasFonspPlutoJl2024,
  title = {Fonsp/{{Pluto}}.Jl: V0.19.42},
  shorttitle = {Fonsp/{{Pluto}}.Jl},
  author = {{Fons van der Plas} and Michiel Dral and Paul Berg and {$\Pi\alpha\nu\alpha\gamma\iota\acute{\omega}\tau\eta\varsigma$} {$\Gamma\varepsilon\omega\rho\gamma\alpha\kappa$}{\'o}{$\pi$}o{$\upsilon\lambda$}o{$\varsigma$} and Rik Huijzer and Miko{\l}aj Boche{\'n}ski and Alberto Mengali and Connor Burns and Benjamin Lungwitz and Hirumal Priyashan and Jerry Ling and Gabriel Wu and Shuhei Kadowaki and Eric Zhang and Felipe S. S. Schneider and Ian Weaver and {Xiu-zhe (Roger) Luo} and Jelmar Gerritsen and Rok Novosel and Supanat and Zachary Moon and Luis M{\"u}ller and Timothy and Vlad Flore and Jeremiah and Ciar{\'a}n O'Mara and Michael Hatherly and {kcin96}},
  year = {2024},
  month = may,
  doi = {10.5281/ZENODO.4792401},
  urldate = {2024-05-24},
  copyright = {Creative Commons Attribution 4.0 International},
  howpublished = {Zenodo}
}

@article{rackauckas2017differentialequations,
  title = {Differentialequations.Jl--a Performant and Feature-Rich Ecosystem for Solving Differential Equations in Julia},
  author = {Rackauckas, Christopher and Nie, Qing},
  year = {2017},
  journal = {Journal of Open Research Software},
  volume = {5},
  number = {1},
  pages = {15},
  publisher = {Ubiquity Press},
  doi = {10.5334/jors.151}
}

@article{vernerNumericallyOptimalRunge2010,
  title = {Numerically Optimal {{Runge}}--{{Kutta}} Pairs with Interpolants},
  author = {Verner, J. H.},
  year = {2010},
  month = mar,
  journal = {Numerical Algorithms},
  volume = {53},
  number = {2},
  pages = {383--396},
  issn = {1572-9265},
  doi = {10.1007/s11075-009-9290-3}
}

@article{danischMakieJlFlexible2021,
  title = {Makie.Jl: {{Flexible}} High-Performance Data Visualization for {{Julia}}},
  author = {Danisch, Simon and Krumbiegel, Julius},
  year = {2021},
  journal = {Journal of Open Source Software},
  volume = {6},
  number = {65},
  pages = {3349},
  publisher = {The Open Journal},
  doi = {10.21105/joss.03349}
}

@misc{myrepo,
    title = {},
    author = {},
    howpublished = "\url{https://github.com/abraemer/Paper_ExploitingEmergentSymmetriesDisorderAveragedDynamics.jl}"
}
\end{document}